# Competing crystallization pathways and cold crystallization kinetics in 10OS5 liquid crystal

Aleksandra Deptuch[1,*], Mirosława D. Ossowska-Chruściel[2], Janusz Chruściel[2], Ewa Juszyńska-Gałązka[1,3]

[1] Institute of Nuclear Physics, Polish Academy of Sciences, Radzikowskiego 152, PL-31342 Kraków, Poland
[2] Faculty of Science, University of Siedlce, 3 Maja 54, Siedlce PL-08110, Poland
[3] Research Center for Thermal and Entropic Science, Graduate School of Science, Osaka University, 560-0043 Osaka, Japan
* corresponding author, aleksandra.deptuch@ifj.edu.pl

## Abstract

The liquid crystalline 4-pentylphenyl-4'-decyloxythiobenzoate is investigated in various temperature programs for determination of crystallization kinetics and glassforming properties. The Avrami model, Augis-Bennett method and isoconversional method are used. Cooling at the 25-30 K/min rate results in formation of the glass of the tilted smectic Y phase with the herring-bone order within layers. Slower cooling leads to the partial or total (2 K/min) crystallization of the metastable Cr2 phase, which during subsequent heating or annealing in a proper temperature transforms to another Cr1 phase. Heating from the vitrified smectic Y leads to cold crystallization of the pure Cr1 phase or the Cr1/Cr2 mix. Both Cr1 and Cr2 are conformationally disordered crystal phases, which is indicated both by the melting entropy values and the dielectric spectra. The results demonstrate that the energy released during cold crystallization can be tuned by thermal history, highlighting 10OS5 as a candidate for thermal energy storage applications.

## 1. Introduction

Cold crystallization is a term referring to the crystallization process that occurs after heating (isothermal) or during heating (non-isothermal) of the material previously supercooled to the metastable state, which can be liquid/amorphous [1-6], liquid crystalline [7,8], or partially disordered crystalline phase [9,10]. Materials which exhibit cold crystallization have a potential use for the energy storage. After cooling below the glass transition temperature $T_g$, they remain in the less energetically favorable state, and they release energy during cold crystallization after they are heated above $T_g$ [5-7]. For application in various conditions, there is a necessity of materials with different temperature ranges of cold crystallization, which is located between $T_g$ and the melting temperature $T_m$. Another feature which needs to be investigated is the kinetics of cold crystallization, i.e., the time it takes and the activation energy of this transition. This paper presents the kinetics of the melt crystallization and cold crystallization of 4-pentylphenyl-4'-decyloxythiobenzoate, $C_5H_{11}$-Ph-S(C=O)-Ph-O-$C_{10}H_{21}$ (Ph – aromatic ring), which is abbreviated as 10OS5 or 10$\bar{S}$5. Previous publications about 10OS5 are focused on its liquid crystalline phases, which include nematic as well as smectics A, C, tilted hexagonal X (G or J) and tilted herring-bone Y (H or K) [11-17]. On cooling at 5 K/min, 10OS5 forms the crystal phase denoted as Cr2, and upon heating at the same rate, there is cold crystallization of the Cr1 phase [17]. In this study, crystallization and Cr2 → Cr1 transition are investigated by differential scanning calorimetry (DSC) and broadband dielectric spectroscopy (BDS). The DSC method enables determination of the crystallization kinetics, temperature range of melt and cold crystallization, and energy released during cold crystallization for different temperature programs. The BDS method is used to check whether the Cr1 and Cr2 phases are characterized by conformational or orientational disorder.

The crystallization process consists of two steps: nucleation and growth of crystallites. Nucleation rate depends on the difference in the free enthalpy $\Delta G$ between the melt and the crystal. It can be approximated as $\Delta G \approx \Delta S_m \Delta T$, were $\Delta S$ is the entropy of melting and $\Delta T = T_m - T$. Thus, $\Delta G$ increases with increasing undercooling $\Delta T$ and nucleation occurs more easily in lower temperatures. The next step, crystal growth, depends on the diffusion rate and because of that, its rate increases with increasing temperature. When the overall rate of crystallization decreases with increasing temperature, it means that crystallization kinetics is restricted mainly by thermodynamic driving force, while the slowing down of crystallization with decreasing temperature means that it is restricted mainly by the diffusion rate [18,19].

Crystallization can be investigated in isothermal conditions, where the sample is cooled or heated to a selected temperature $T_{cr}$ and kept in this temperature until crystallization is finished. Another method is study in non-isothermal conditions, where the sample is cooled or heated with a selected constant rate $\phi$ during crystallization. Based on the DSC thermograms, the relative crystallization degree can be determined by integration of the exothermic anomaly related to crystallization over time:

$$x(t) = \frac{\int_{t_0}^{t} \Phi(t)dt}{\int_{t_0}^{t_{end}} \Phi(t)dt}, \quad (1)$$

where $\Phi(t)$ is the heat flow, $t_0$ is the initialization time, and $t_{end}$ is the finish time of crystallization. For the non-isothermal crystallization, the integration is performed over temperature. As crystallization may be incomplete in some conditions, e.g., during fast cooling, for calculation of the relative conversion degree one takes as 100% the amount of the crystal phase formed during a given process, while the absolute degree of crystallization can be below 100% at the same time [19,20]. To distinguish these parameters, the relative degree of crystallization is denoted here by $x(t)$ and absolute degree of crystallization by capital $X(t)$. For the complete crystallization, $x(t) = X(t)$.

The Avrami model is widely used to describe the $x(t)$ dependence:

$$x(t) = 1 - \exp\left(-\left(\frac{t-t_0}{\tau_{cr}}\right)^n\right), \quad (2)$$

where $\tau_{cr}$ is the characteristic crystallization time and $n$ is the Avrami parameter dependent on the nucleation rate and shape of crystallites. In conditions of constant amount of nuclei, $n$ = 1, 2, 3, 5 for needle-like, plate-like, isotropic, and sheaf-like growth of crystallites, while for constant nucleation rate the respective values are $n$ = 2, 3, 4, 6. It should be noted that the Avrami exponent is an effective parameter reflecting combined contributions from nucleation, growth dimensionality, and impingement effects [21-23]. The activation energy $E_{cr}$ of crystallization can be obtained from the Arrhenius plot of $\ln \tau_{cr}$ vs. $1/T_{cr}$, which is expected to show a linear dependence with a slope equal to $E_{cr}/R$, where $R$ is the gas constant [24,25].

There are various methods to determine the activation energy of non-isothermal crystallization. The Augis-Bennett model [26] uses the onset $T_o$ and peak $T_p$ temperatures of the exothermic anomaly related to crystallization:

$$\ln\left(\frac{\phi}{T_p - T_o}\right) = C - \frac{E_{cr}}{RT_p}, \tag{4}$$

where $C$ is the fitting parameter. The earlier Kissinger model, which was the basis for the Augis-Bennett model, was derived for $n$ = 1 only [27]. Despite that, both methods often give comparable activation energies [2,25,28], but sometimes larger discrepancy occurs [29] because the Kissinger model is less general. Using the $E_{cr}$ values and the full-width at half-height of the exothermic anomaly $\Delta T_{FWHM}$, the Augis-Bennett model enables determination of the Avrami parameter as [26]:

$$n = \frac{2.5RT_p^2}{E_{cr}\Delta T_{FWHM}}. \tag{5}$$

The isoconversional method enables determination of the activation energy as a function of the relative degree of conversion and temperature [20,30-33] using the formula:

$$\frac{dx(t)}{dt} = f(x)A\exp\left(-\frac{E_{cr}}{RT}\right), \tag{6}$$

where $f(x)$ is the reaction model and $A$ is the pre-exponential constant [30,31]. The $E_{cr}$ value for a selected value of $x$ is obtained from the slope of the plot of $\ln(dx/dt)$ vs. $1000/T_x$, where both the conversion rate $dx/dt$ and temperature $T_x$ correspond to a moment where a given $x$ was reached.

Despite extensive studies on the phase behavior of 10OS5, the interplay between melt crystallization, cold crystallization, and the metastability of intermediate phases (SmY and Cr2) has not been systematically investigated, particularly in the context of energy storage applications.

## 2. Experimental and computational details

Synthesis of 4-pentylphenyl-4'-decyloxythiobenzoate was performed according to the procedure described elsewhere [34].

The sample for the DSC measurements weighted 2.80 mg and was placed in an aluminum pan. The DSC thermograms were collected with the DSC 2500 calorimeter in the temperature range of 173-373 K with various cooling/heating rates of 2-30 K/min or in the isothermal conditions in selected temperatures. The DSC results were analyzed in TRIOS and OriginPro.

The samples for the BDS experiment were placed between two gold electrodes with polytetrafluoroethylene spacers and had a thickness of 75 μm. The BDS spectra were collected with the Novocontrol Technologies spectrometer in the temperature range of 173-373 K and in the frequency range of 0.1-$10^7$ Hz. The spectra were fitted in OriginPro with the complex model of the dielectric permittivity, described in Section 3.

The DFT calculations for an isolated molecule in Gaussian 16 [35] were performed using the B3LYP-D3(BJ) exchange-correlation functional [36-39] and def2TZVPP basis set [40]. The Avogadro program [41] was used to prepare the starting model of a molecule and to generate image of the optimized model.

## 3. Results and discussion

### *3.1. Non-isothermal melt crystallization*

The 10OS5 compound shows five mesophases during cooling at 2-30 K/min (Figure 1). The nematic and SmA phase are enantiotropic phases, as they appear both during cooling and heating. Three monotropic mesophases, SmC, SmX and SmY, are present only during cooling. The phase transition temperatures between mesophases depend weakly on the cooling rate and the transition temperatures in kelvins, determined by the linear extrapolation to 0 K/min, are as follows: Iso (358.0) N (352.4) SmA (334.3) SmC (319.5) SmX (312.0) SmY on cooling and SmA (352.2) N (358.0) Iso on heating. On further cooling, the sample's behavior becomes strongly dependent on the cooling rate. For 2 K/min, the onset temperature $T_o$ of the melt crystallization is 302 K and the peak temperature $T_p$, corresponding to the maximal crystallization rate, is 298 K. Both $T_o$ and $T_p$ shift to lower temperatures on faster cooling. Also, the enthalpy changes $\Delta H_{cr}$ related to crystallization decrease with an increasing cooling rate, from 13.5 kJ/mol at 2 K/min to only 0.4 kJ/mol at 20 K/min, which indicates an increasing fraction of remaining SmY. For 25-30 K/min, the crystallization peak is not visible and the sample apparently consists only of supercooled SmY. During heating with the same rate as in the preceding cooling run, cold crystallization is observed (inset in Figure 1b). For 2-3 K/min, the exothermic anomaly has an onset temperature of 294 K, for 4-10 K/min, $T_o$ decreases with an increasing heating rate from 290 to 266 K, and for 12-30 K/min, $T_o$ increases gradually to 278 K. By comparison of the cooling and heating runs, it can be concluded that for 2-3 K/min, the SmY → Cr2 transition is completed during cooling and cold crystallization which occurs during heating is the Cr2 → Cr1 transition. For 4-10 K/min, the SmY → Cr2 transition is only partial and during heating one observes the completion of the SmY → Cr2 transition overlapped with Cr2 → Cr1 at higher temperatures. For 12-30 K/min, the fraction of Cr1 formed during cooling is small or even absent (25-30 K/min), therefore, the SmY → Cr2 and Cr2 → Cr1 transitions contribute to similar extent to the exothermic anomaly observed during heating. Upon further heating, the endothermic anomaly from the melting of Cr1, with $T_o$ = 336 K, is visible in all thermograms, which means that after cold crystallization, the fraction of this phase is dominant. For 25-30 K/min, there is also a well-visible endothermic anomaly with $T_o$ = 330 K, corresponding to the melting of Cr2, which means that for fast heating, the Cr2 → Cr1 transition is incomplete. A remaining fraction of Cr2 is seen also for 3-6 K/min, but the area of this anomaly is very small ($\Delta H$ = 0.2-0.6 kJ/mol), thus, the Cr2 → Cr1 transition is almost completed.

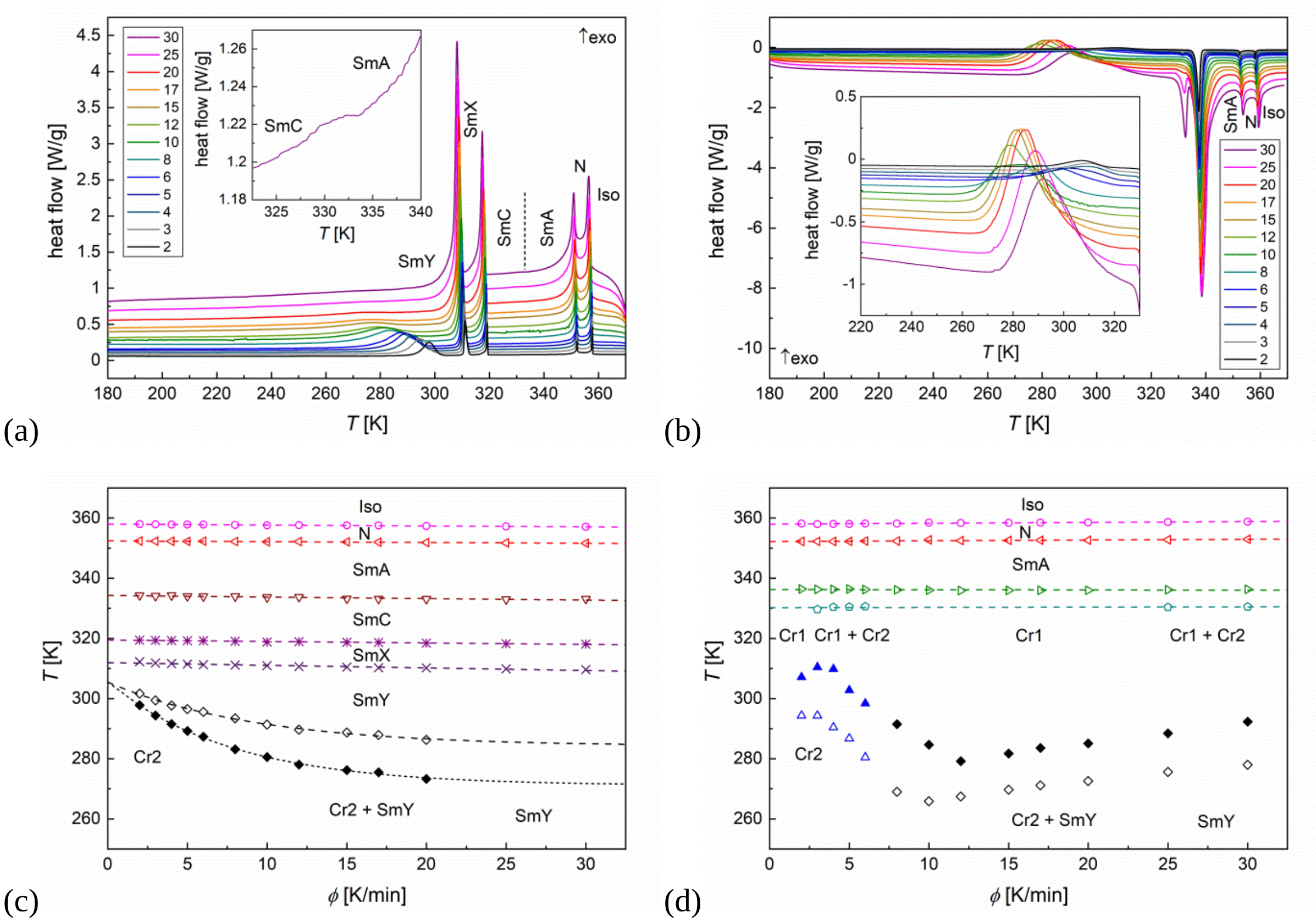

Figure 1. DSC thermograms of 10OS5 collected during cooling at various rates (a) and heating at the same rate (b), and the phase transition temperatures determined by DSC for cooling (c) and heating (d). The cooling/heating rates in K/min are given in legends. The insets in (a) and (b) show the close-ups to the SmA → SmC transition occurring at 30 K/min and cold crystallization at various heating rates, respectively.

The analysis by the Augis-Bennet method (Equation 4, Figure 2a) indicates two regimes of non-isothermal melt crystallization with different values of $E_{cr}$: −23(2) kJ/mol for slow cooling at 2-6 K/min and −45(4) kJ/mol for fast cooling at 8-20 K/min. By application of Equation (5), the Avrami parameter $n$ = 2.3-3.0, corresponding to mainly 2-dimensional crystal growth, is obtained for 8-20 K/min. Meanwhile, for 2-6 K/min, one obtains unphysical $n$ = 8-19. It is caused by small absolute $E_{cr}$ value, because in their derivation, Augis and Bennett made an assumption that $E_{cr}$ differed significantly from zero [26]. The isoconversional method (Equation 6, Figure 2b,c), which is applied here to almost the whole range of $x(T)$, is the most versatile and it can be used for various $E_{cr}$ values. Different slopes are obtained for 3-8 K/min and 10-20 K/min rates, corresponding to $E_{cr}$ = −(9-24) kJ/mol and –(61-178) kJ/mol, respectively. The negative apparent activation energy indicates that the overall crystallization rate is governed by the thermodynamic driving force (undercooling), rather than molecular mobility. Such behavior reflects the competition between nucleation and diffusion processes. The difference between $E_{cr}$ obtained by the isoconversional method and Augis-Bennett method increases with increasing $x$. For slow cooling, the absolute $E_{cr}$ value decreases with increasing $x$, indicating the growing influence of the diffusion rate. For fast cooling, absolute $E_{cr}$ increases, especially for later stages of crystallization, which means probably that the sample is cooled down below the optimal temperature range for nucleation. This suggests that the system is driven outside the optimal temperature window for nucleation, leading to a decrease in nucleation efficiency despite increasing undercooling. The results for 2 K/min deviate from the linear dependence obtained for 3-8 K/min, which means that for very slow cooling, there can be another crystallization regime.

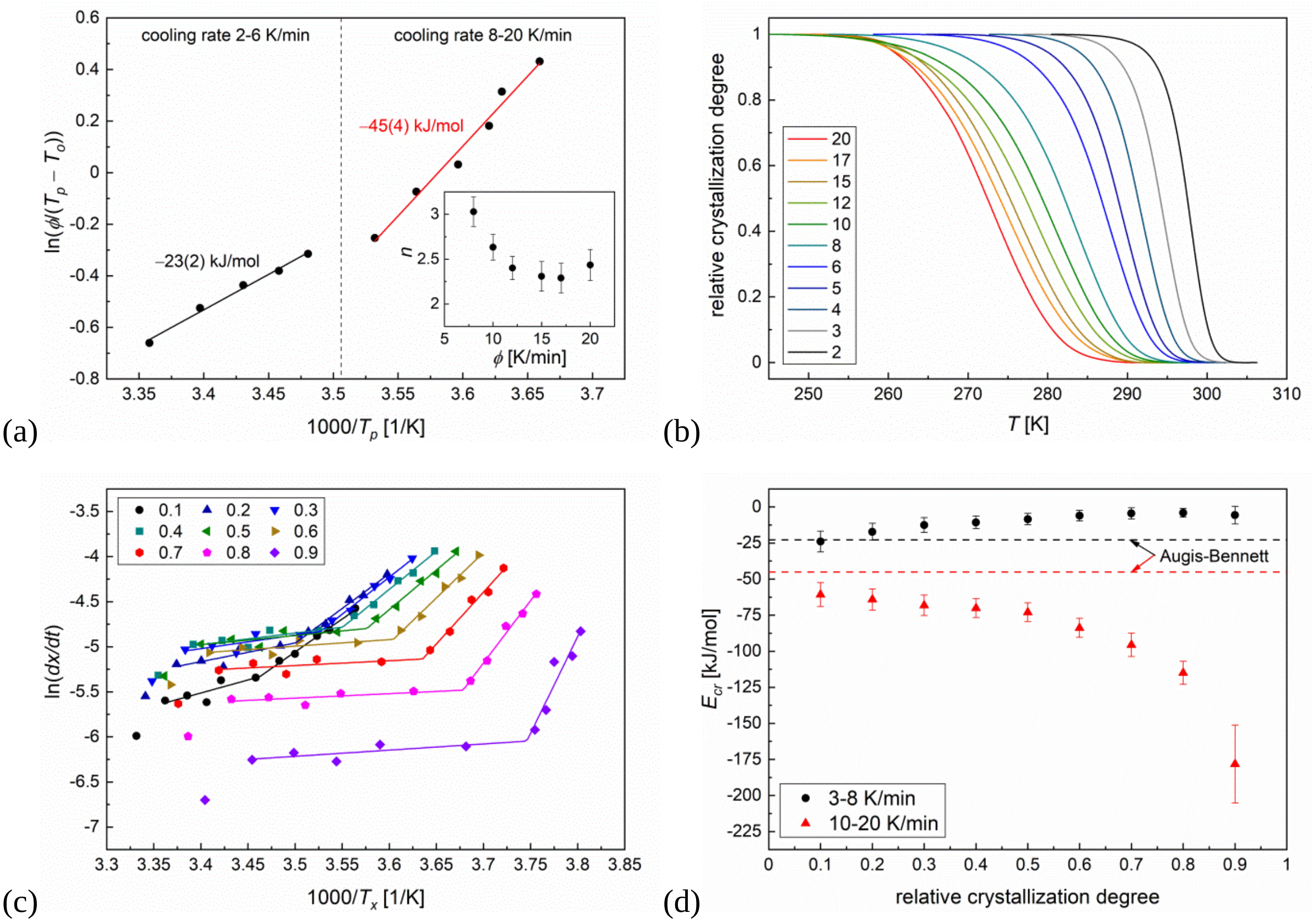

Figure 2. The effective activation energy of non-isothermal melt crystallization obtained with the Augis-Bennett method (a), relative degree of non-isothermal melt crystallization for various cooling rates given in K/min in a legend (b), determination of the effective activation energy by the isoconversional method (c), and comparison of the effective activation energy obtained by both methods (d). The inset in (a) shows the Avrami exponent vs. cooling rate.

### *3.2. Isothermal melt crystallization*

Melt crystallization in isothermal conditions was investigated by cooling the sample at 20 K/min from 373 K to a selected crystallization temperature $T_{cr}$ = 290-300 K, where the DSC thermograms as a function of time were collected (Figure 3a). Two exothermic anomalies are observed, the first corresponding to the SmY → Cr2 transition and another, wider, corresponding to the Cr2 → Cr1 transition. For $T_{cr}$ = 300 K, only the first anomaly is present. The endothermic peaks observed upon subsequent heating (Figure 3b) show that for $T_{cr}$ = 290-294 K, mainly the Cr1 phase, with small remains of Cr2, is present in the sample after crystallization. For $T_{cr}$ = 296-298 K, the amount of Cr2 increases, but most of the sample consist of Cr1, and for $T_{cr}$ = 300 K, the Cr1 phase is not formed. The $x(t)$ dependences both for SmY → Cr2 (Figure 3c) and Cr2 → Cr1 (Figure 3d) are fitted with the Avrami model (Equation 2, Table 1), which describes well the experimental results, except for the initial stage, which likely corresponds to an induction period not captured by the Avrami model. Thus, the initialization time $t_0$ presented in Table 1 corresponds to the main crystallization process following the Avrami description, while overall crystallization begins earlier. The characteristic crystallization time $\tau_{cr}$ is equal to 94-112 s for SmY → Cr2, while for Cr2 → Cr1, it is more than one order of magnitude larger, 980-5500 s. The activation plots of $\tau_{cr}$ have both negative, but significantly different slopes, with corresponding $E_{cr}$ = −7(1) kJ/mol for SmY → Cr2 and $E_{cr}$ = −162(8) kJ/mol for Cr2 → Cr1. The significantly higher magnitude of the activation energy for the Cr2 → Cr1 transition suggests a stronger dependence on thermodynamic stability differences between crystal phases compared to nucleation from the smectic phase. The Avrami parameter $n$ = 1.5-1.7 corresponds to 1-dimensional growth of Cr2 and $n$ = 2.5-2.8 means that the growth of Cr1 is 2-dimensional. Only for $T_{cr}$ = 298 K, $n$ = 1.7 indicates rather 1-dimensional growth of Cr1.

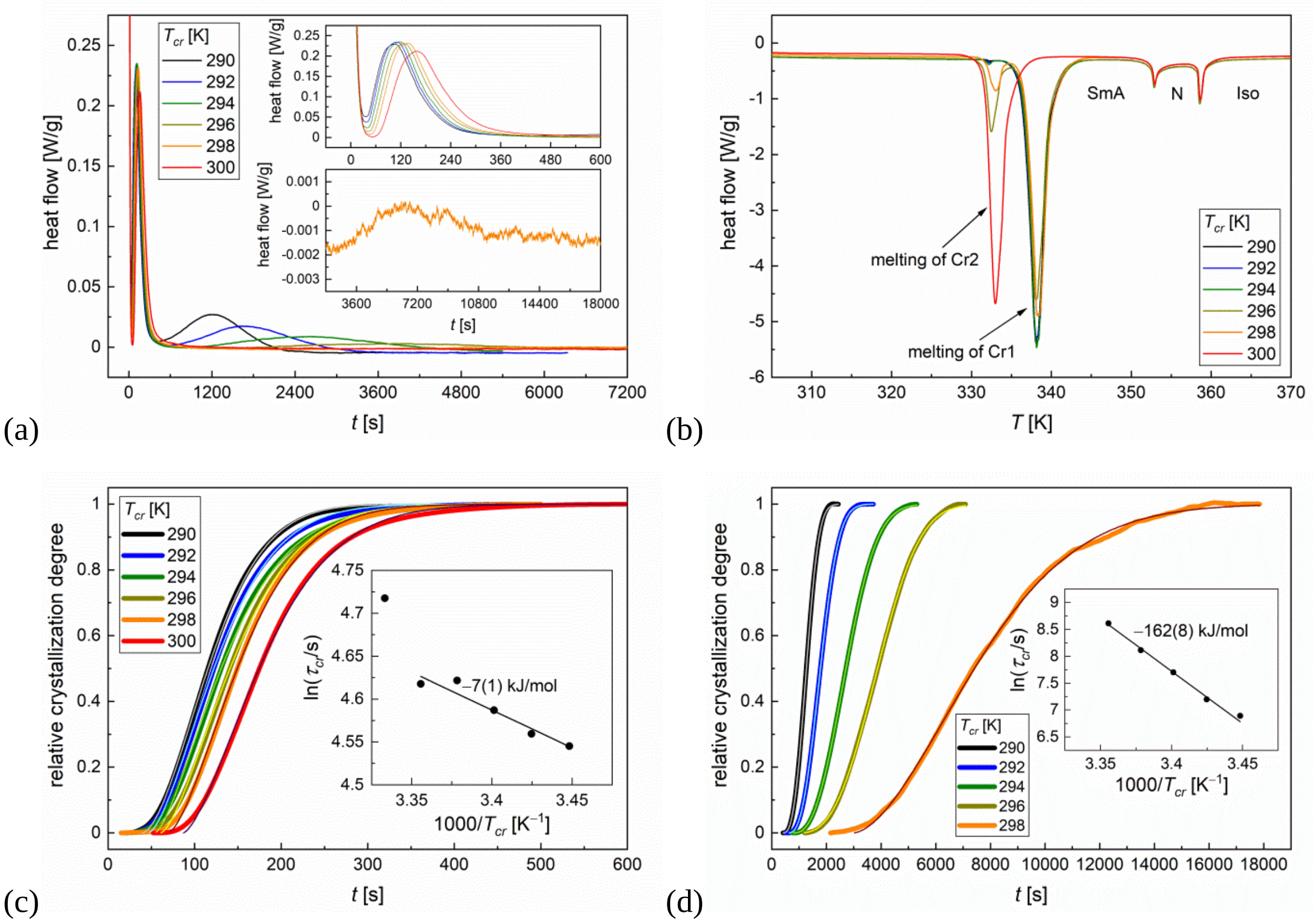

Figure 3. DSC thermograms of 10OS5 collected during isothermal melt crystallization (a) and during heating at 10 K/min (b), and relative crystallization degree of the SmY → Cr2 (c) and Cr2 → Cr1 (d) transitions with fitting results of the Avrami model. The insets in (c) and (d) are the corresponding activation plots of the characteristic crystallization time.

Table 1. Parameters of the Avrami model determined by fitting Equation (2) to the relative crystallization degree of isothermal melt crystallization of 10OS5, investigated by DSC.

| transition | SmY → Cr2 | | | Cr2 → Cr1 | | |
|---|---|---|---|---|---|---|
| $T_{cr}$ [K] | $t_0$ [s] | $\tau_{cr}$ [s] | $n$ | $t_0$ [s] | $\tau_{cr}$ [s] | $n$ |
| 290 | 39.3(2) | 94.2(2) | 1.72(1) | 405(1) | 983(1) | 2.78(1) |
| 292 | 46.7(2) | 95.5(2) | 1.56(1) | 622(1) | 1340(1) | 2.54(1) |
| 294 | 54.4(2) | 98.2(2) | 1.45(1) | 805(1) | 2214(1) | 2.47(1) |
| 296 | 60.9(2) | 101.7(2) | 1.52(1) | 950(2) | 3334(2) | 2.76(1) |
| 298 | 69.4(2) | 101.3(2) | 1.54(1) | 2973(2) | 5493(2) | 1.72(1) |
| 300 | 86.8(2) | 111.9(2) | 1.51(1) | - | - | - |

### *3.3. Non-isothermal Cr2 → Cr1 transition*

To observe the Cr2 → Cr1 transition for various heating rates, the sample was cooled from isotropic liquid to 290 K and kept at this temperature for 6.5 min, which is enough time for completion of the SmY → Cr2 transition, but before the start of the Cr2 → Cr1 transition. The sample prepared in such a way was then heated at 2-30 K/min (Figure 4a). The DSC thermograms show that the transition to Cr1 occurs slowly above 290 K and only for the 2, 3, 4 K/min heating rates, the exothermic anomaly from the Cr2 → Cr1 transition is clearly visible above the baseline. Thus, the kinetic of this transition could have not been investigated. The fraction of the Cr2 and Cr1 phases upon melting vs. the heating rate (Figure 4b) was determined based on the area of two endothermic peaks at 331 K and 336 K. While for 2 K/min, the Cr2 → Cr1 transition is almost complete, the fraction of Cr1 decreases significantly with the increasing heating rate and for 15-30 K/min, the Cr1 fraction is only ca. 10%.

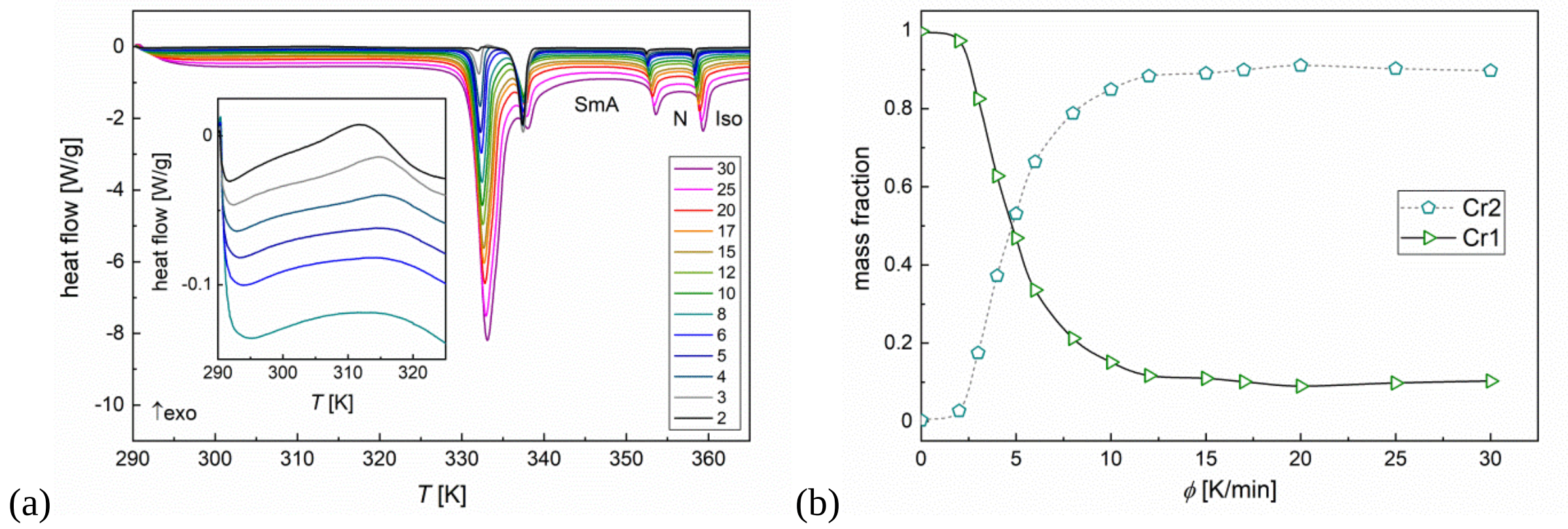

Figure 4. DSC thermograms of 10OS5 collected during heating after isothermal melt crystallization at 290 K (a) and fractions of crystal phases upon melting vs. the heating rate (b). The values for 0 K/min correspond to the isothermal Cr2 → Cr1 transition in 290 K, investigated in Section 3.2.

### *3.4. Non-isothermal cold crystallization*

The kinetics of cold crystallization was investigated for the sample cooled down at 30 K/min to the vitrified SmY phase (Figures 5 and 6). During heating at 2-30 K/min, cold crystallization is always observed. For 2-12 K/min, both the SmY → Cr2 and Cr2 → Cr1 transitions are completed, while for 15-30 K/min, the fraction of the remaining Cr2 phase gradually increases with an increasing heating rate. Due to significant overlap between the SmY → Cr2 and Cr2 → Cr1 transitions, the exothermic signal was treated as a single effective process. While this approach simplifies the analysis, it should be noted that the extracted kinetic parameters represent averaged contributions from both transformations. The Augis-Bennett method (Figure 6a) indicates the same activation energy in the 2-30 K/min range, $E_{cr}$ = 38.7(9) kJ/mol. The Avrami exponent $n$ decreases with the increasing heating rate. For 2-5 K/min, the $n$ values indicate partially isotropic, partially sheaf-like growth, for 6-12 K/min mainly isotropic growth and for 15-30 K/min mainly plate-like growth. The isoconversional analysis (Figure 6c) shows the same activation energy for a given $x$ for 5-30 K/min, while the points for 2-4 K/min deviate from the linear dependence (the SmY → Cr2 and Cr2 → Cr1 transitions are better separated for slow heating, which distorts the $x(T)$ dependences). The activation energy decreases slowly during crystallization until $x$ = 0.5, where it is in the best agreement with the value obtained by the Augis-Bennett method, and in the later stages it significantly increases. The positive apparent activation energy indicates that the crystallization kinetics is controlled primarily by molecular mobility (diffusion-limited regime).

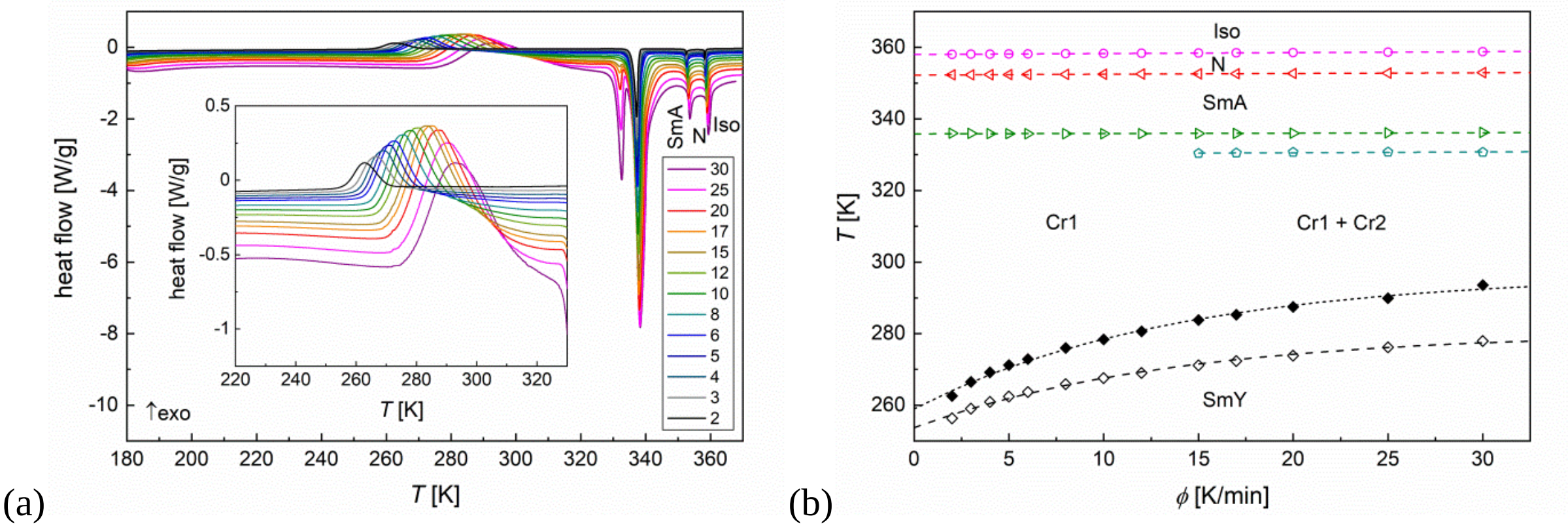

Figure 5. DSC thermograms of 10OS5 collected during heating with various rates after cooling at 30 K/min (a) and the phase transition temperatures on heating (b).

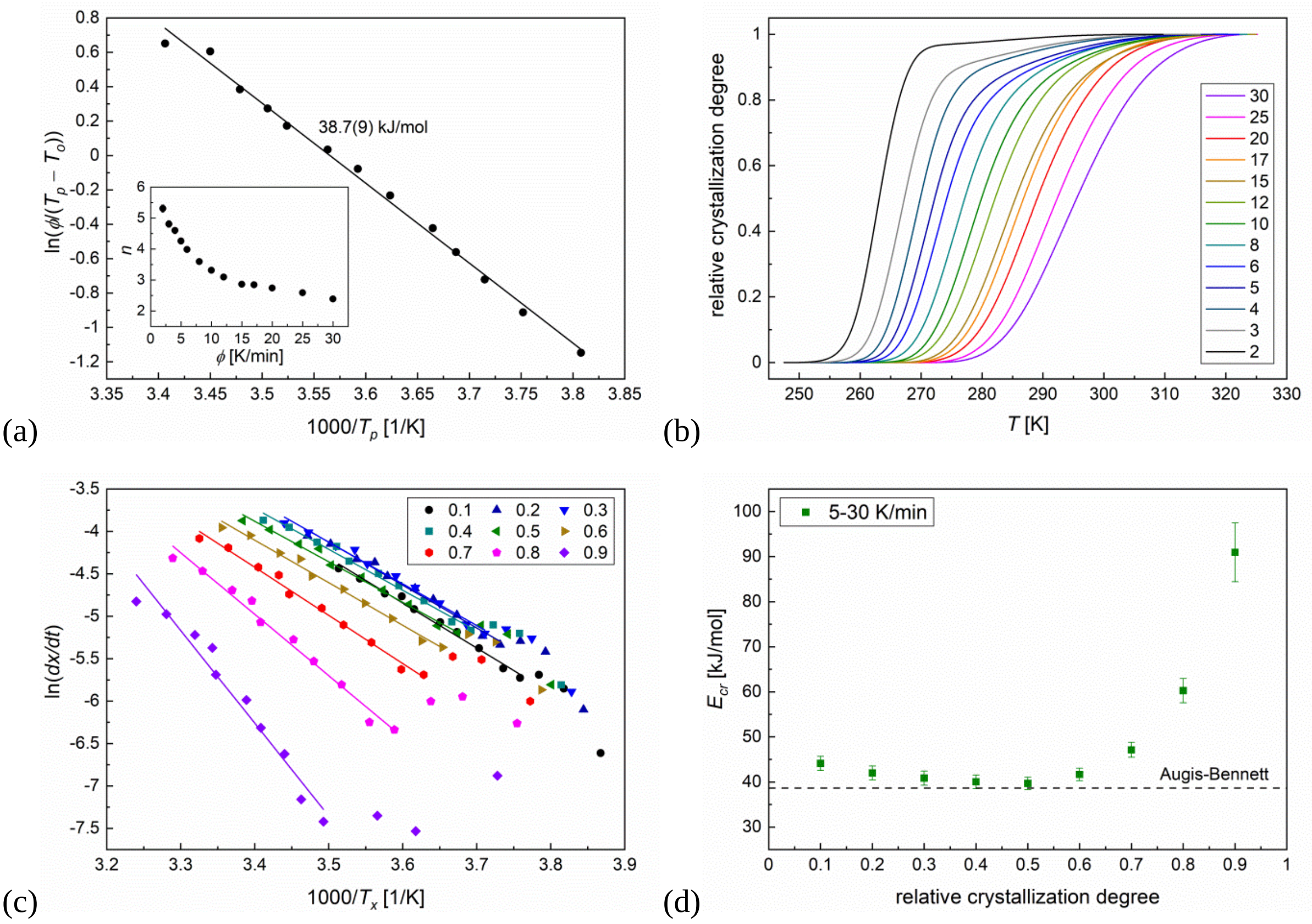

Figure 6. The effective activation energy of non-isothermal cold crystallization obtained with the Augis-Bennett method (a), relative degree of non-isothermal cold crystallization of 10OS5 for various cooling rates given in K/min in a legend (b), determination of the effective activation energy by the isoconversional method (c), and comparison of the effective activation energies obtained by both methods (d). The inset in (a) shows the Avrami exponent vs. the heating rate.

### *3.5. Isothermal cold crystallization*

To investigate cold crystallization in isothermal conditions, the sample was heated to 373 K, cooled down to 173 K at 30 K/min and heated with the same rate to $T_{cr}$ = 248, 251, 253, 256, or 258 K, where isothermal crystallization occurred (Figure 7a). After cold crystallization at each $T_{cr}$, the sample was heated to 373 K at 10 K/min to observe melting of the crystal phase (inset in Figure 7a). The melting temperature $T_m$ = 336 K after cold crystallization in each $T_{cr}$ indicates that only the Cr1 phase is formed, as the Cr2 → Cr1 transition is not visible between $T_{cr}$ and $T_m$. The crystallization degree was fitted with the Avrami model (Equation 2, Figure 7b). The obtained parameters are presented in Table 2. The growth of crystallites is rather 1-dimensional, as indicated by low $n$ = 1.6-1.7. The characteristic crystallization time is shorter for higher $T_{cr}$ and changes according to the Arrhenius formula (inset in Figure 7b). The activation energy is positive and equal to 86(2) kJ/mol, thus, isothermal cold crystallization of 10OS5 is restricted by the diffusion rate.

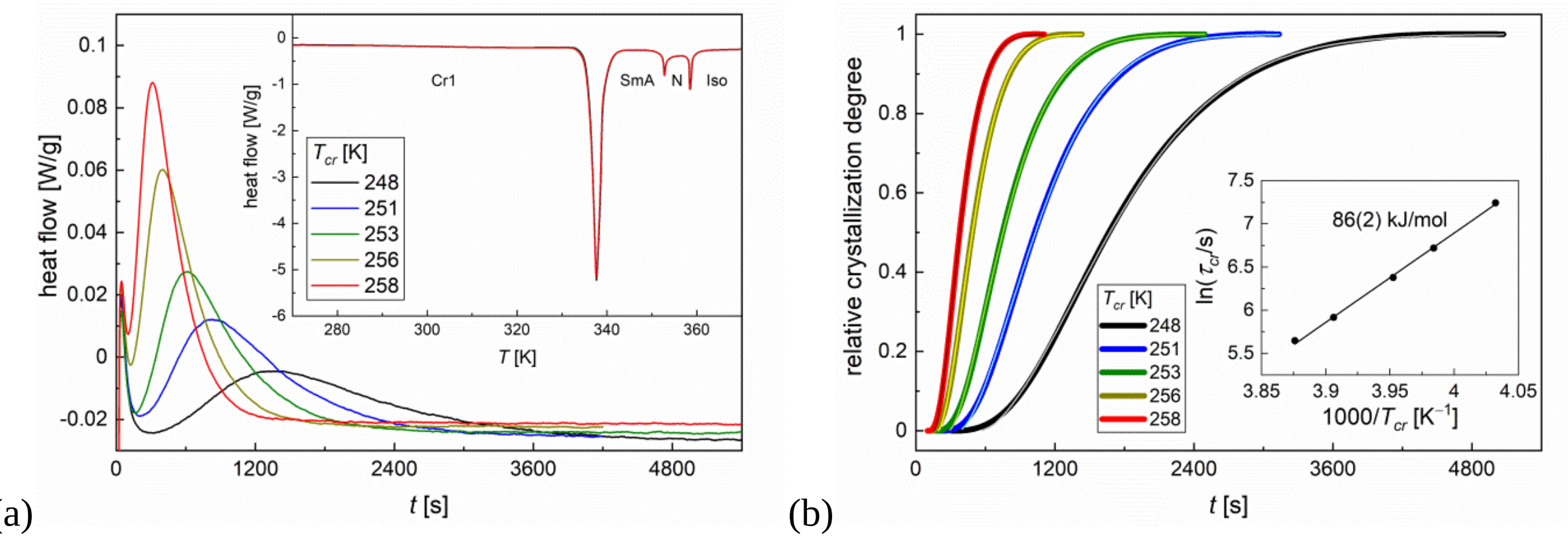

Figure 7. DSC thermograms of 10OS5 collected during the isothermal cold crystallization (a) and the corresponding crystallization degree (b). Inset in (a) shows the DSC thermograms collected during heating at 10 K/min after isothermal cold crystallization. Inset in (b) is the activation plot of the characteristic time of isothermal cold crystallization.

Table 2. Parameters of the Avrami model determined by fitting Equation (2) to the relative crystallization degree of isothermal cold crystallization of 10OS5, investigated by DSC.

| transition | SmY → Cr1 | | |
|---|---|---|---|
| $T_{cr}$ [K] | $t_0$ [s] | $\tau_{cr}$ [s] | $n$ |
| 248 | 555.2(4) | 1395.4(5) | 1.66(1) |
| 251 | 370.4(3) | 828.1(3) | 1.61(1) |
| 253 | 285.3(3) | 589.6(3) | 1.56(1) |
| 256 | 190.2(2) | 371.4(2) | 1.66(1) |
| 258 | 151.9(2) | 283.9(2) | 1.67(1) |

### *3.6. Dielectric relaxation processes*

The result of BDS measurement is the complex dielectric permittivity $\varepsilon^*$ as a function of frequency $f$ of a weak external electric field. The real and imaginary parts of $\varepsilon^*$ are the dielectric dispersion $\varepsilon'$ and absorption $\varepsilon''$. The model used in analysis of the results for 10OS5 is the Cole-Cole model [42] including the contribution of the conductivity [43]:

$$\varepsilon^*(f) = \varepsilon'(f) - i\varepsilon''(f) = \varepsilon_\infty + \sum_j \frac{\Delta\varepsilon_j}{1+(2\pi i f\tau_j)^{1-a_j}} - \frac{is}{(2\pi f)^m}, \quad (7)$$

where $\varepsilon_\infty$ is the dielectric dispersion at the high frequency limit, $\Delta\varepsilon$ is the dielectric strength, $\tau$ is the relaxation time, $a$ is the Cole-Cole parameter (for $a$ = 0 one obtains the Debye model), $s$ and $m$ are parameters describing the conductivity contribution at low frequencies. If $m$ = 1, then $s = \sigma/\varepsilon_0$, where $\sigma$ is the ionic conductivity and $\varepsilon_0$ is the vacuum permittivity. Three temperature programs were applied: (1) measurements on slow cooling from 373 to 173 K and on heating to 373 K, (2) direct cooling at 2 K/min from 373 to 173 K and measurement on heating to 373 K, (3) direct cooling at 15 K/min from 373 to 173 K and measurement on heating to 373 K. The representative BDS spectra are presented in Figure 8 and the activation plot of relaxation times is shown in Figure 9.

The relaxation time of the process observed in the Iso, N, SmA, and SmC phases (Figure 8a) decreases with increasing temperature according to the Arrhenius formula. This process is interpreted as the molecular s-process, corresponding to rotations of molecules around their short axes [13,43-45]. The activation energy of the s-process is sensitive to the Iso → N and N → SmA transitions and is equal to 85(13), 254(8) and 91.5(4) kJ/mol in the Iso, N and SmA/SmC phases, respectively. The activation energy $E_a$ given in [13] for N and SmA/SmC is equal to 206 and 106 kJ/mol. Differences with results obtained in this study may arise from a smaller number of experimental points used in [13] (3 points for N and 5 points for SmA/SmC). Both the present study and [13] indicate an increased activation energy of the s-process in the N phase. Such effect was observed also for, e.g., the nCB series [44]. The arise of the molecular tilt at the SmA/SmC transition does not affect $E_a$ of the s-process. In the tilted hexagonal SmX phase, there are two relaxation processes (Figure 8b). The low-frequency process is the s-process, with an activation energy of 133.9(3) kJ/mol. Its relaxation time is shifted to lower values at the SmC → SmX transition, which is caused by appearance of the hexagonal order [43,44]. The high-frequency process is the s-process from the remaining SmC phase and it seems to show an inverse Arrhenius behavior, i.e., decrease of the relaxation time on cooling. However, it is likely the effect of the decreasing domains of SmC. In the crystal phases (Figure 8c), the polarization or Maxwell-Wagner-Sillars process [46] is noticeable at low frequencies. It appears also above the melting temperature in spectra collected upon heating, but it is not the subject of this work and is not discussed. According to results for isothermal melt crystallization (Figure 3a), the Cr2 → Cr1 transition occurs during gradual slow cooling in the BDS experiment, thus, below the room temperature the sample is mainly in the Cr1 phase. In lower temperatures, two relaxation processes appear: the low-frequency process $\mathrm{cr1_{LF}}$ with a non-Arrhenius behavior and the high-frequency process $\mathrm{cr1_{HF}}$ with an Arrhenius behavior and an activation energy of 72(4) kJ/mol. The $\mathrm{cr1_{LF}}$ process leaves the investigated frequency range below 237 K. The $\mathrm{cr1_{HF}}$ process is still visible below 237 K, but it is very weak and overlaps with the tail of $\mathrm{cr1_{LF}}$, therefore, it was not fitted. The relaxation time of the $\mathrm{cr1_{LF}}$ process is described by the Vogel-Fulcher-Tammann equation [43,47]:

$$\tau(T) = \tau_0 \exp\left(\frac{B}{T-T_V}\right), \tag{8}$$

where $\tau_0$ is the pre-exponential factor ($\log_{10}(\tau_0/s) = -9.1(2)$), $T_V$ = 211(2) K is the Vogel temperature at which the relaxation time diverges, and $B$ = 424(39) K describes the slope of $\ln \tau(T)$ plotted against $(T - T_V)^{-1}$ (if $T_V$ = 0, then $B = E_a/R$). Such behavior is typical for the α-process, observed for glassformers [3,10,47,48], which indicates the occurrence of the glass transition in the Cr1 phase. The glass transition temperature, which corresponds to $\tau$ = 100 s [47], is equal to $T_g$ = 228.0(8) K. The deviation of the $\tau(T)$ dependence from the Arrhenius formula is described by the fragility parameter $m_f$ [47]:

$$m_f = \left.\frac{d \log_{10} \tau(T)}{d(T_g/T)}\right|_{T=T_g} = \frac{BT_g}{\ln 10 (T_g - T_V)^2}. \tag{9}$$

For 10OS5, $m_f$ = 153(10). Such high value indicates the fragile glassformer. The glass transition in the crystal phases may be the effect of the conformational disorder (CONDIS phase) or orientational disorder (ODIC phase) [10,29,48]. The entropy changes corresponding to the positional, orientational, and conformational degrees of freedom in a crystal phase are equal to 7-14, 20-50, and $k$(7-12) J/(mol·K), respectively, where $k$ is the number of rotatable parts of a molecule [49]. In the 10OS5 molecule, $k$ = 19 (number of torsional angles which can be reasonably changed) and with an assumption of the perfectly ordered crystal, the expected entropy of melting would be at least $\Delta S_m$ = 160 J/(mol·K). Meanwhile, the DSC results give $\Delta S_m$ = 109 J/(mol·K) for Cr1 and 88 J/(mol·K) for Cr2, therefore, some disorder is expected in both crystal phases.

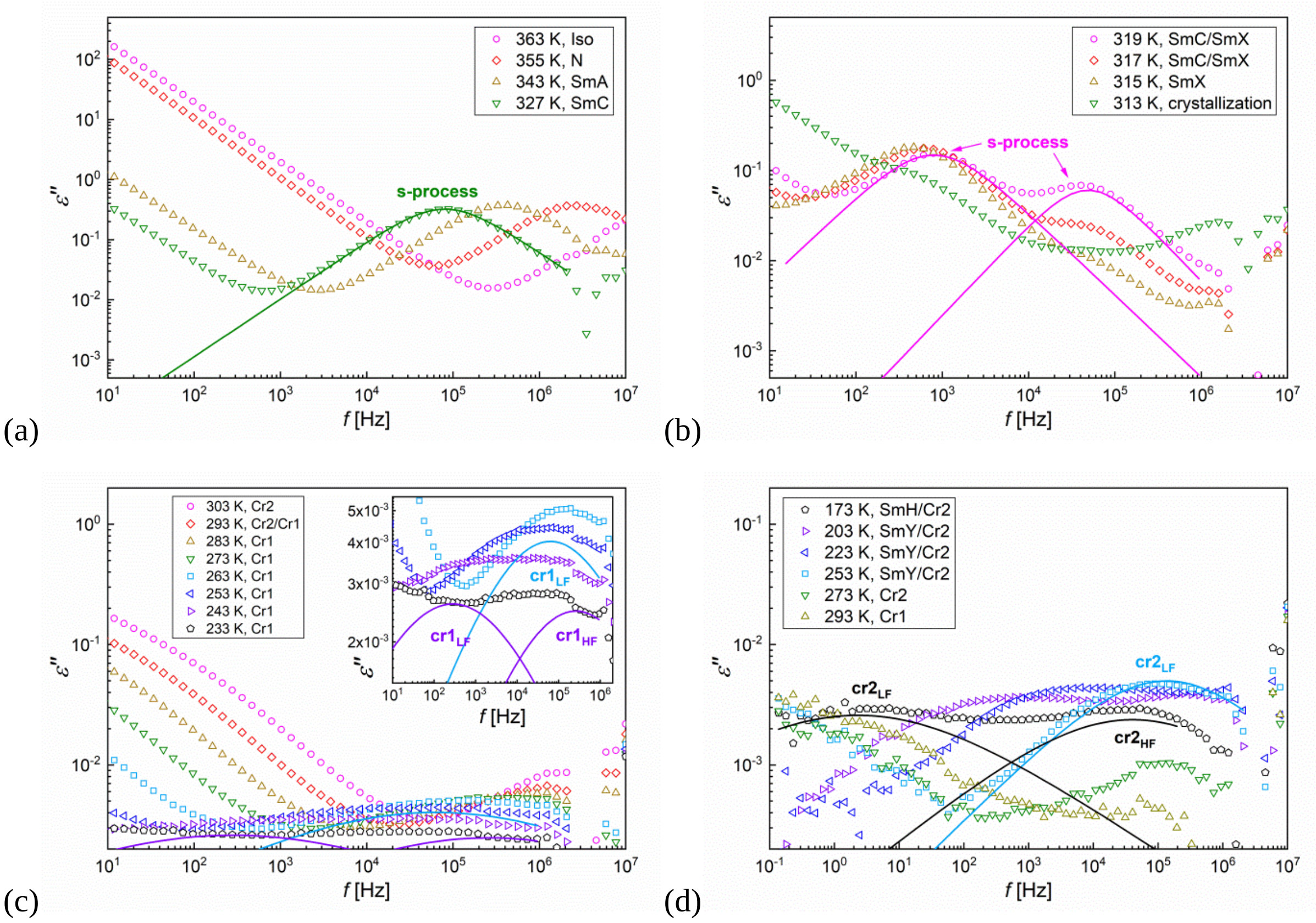

Figure 8. Exemplary dielectric absorption spectra of 10OS5 collected on slow cooling (a-c) and upon heating after cooling at 15 K/min (d). The lines show the fitting results of the (7) formula, with omitted conductivity contribution. The inset in (c) is the enlarged part of the main panel.

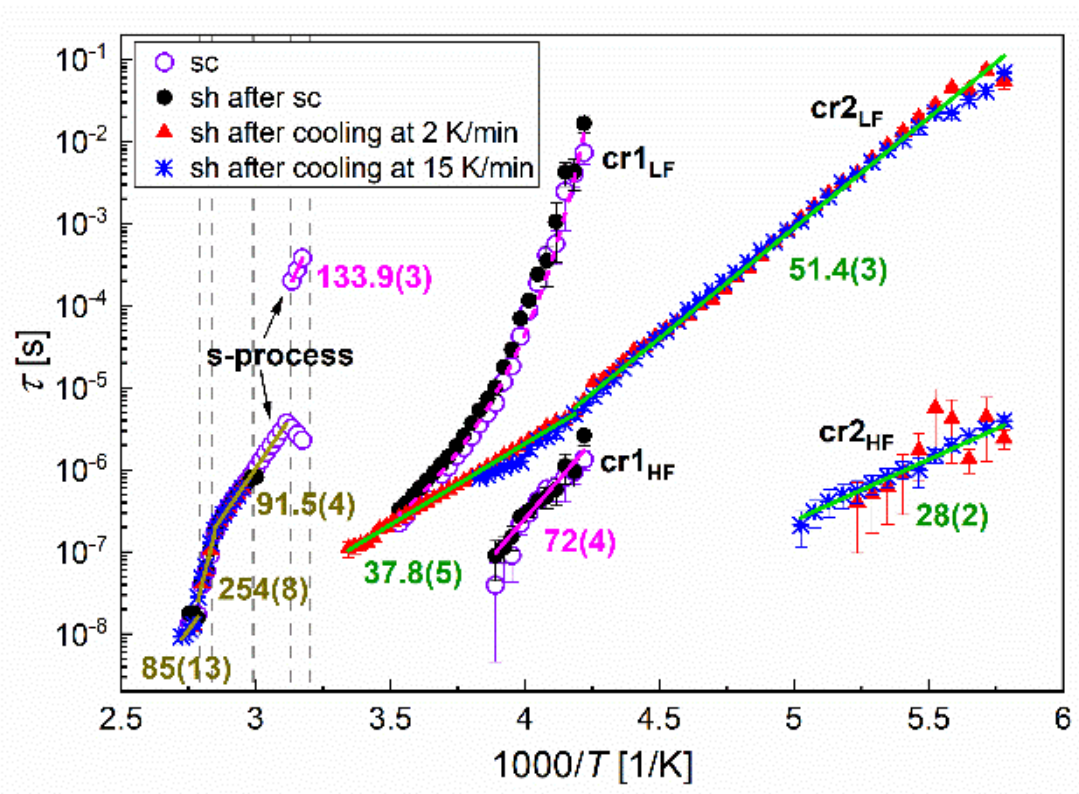

Figure 9. Activation plot of the relaxation times obtained from the BDS spectra of 10OS5 (sc – slow cooling, sh – slow heating). The activation energies are in kJ/mol.

The BDS spectra collected upon heating after cooling at 2 and 15 K/min reveal different relaxation processes than observed after slow gradual cooling (Figure 8d and Figure 9). The absolute degree of the SmY → Cr2 transition can be determined by comparison of the corresponding enthalpy changes with the maximal enthalpy change of 13.5 kJ/mol during isothermal melt crystallization in 300 K. The enthalpy changes of the SmY → Cr2 transition during cooling at 2 and 15 K/min are 13.5 and 5.6 kJ/mol, which indicates the absolute crystallization degree of 100% and 12%, respectively. The surprising observation is that both for pure Cr2 and SmY/Cr2 mix, there are the same relaxation processes, and no additional processes from the remaining SmY phase are detected in the former case. The previous XRD results show that the Cr2 phase is lamellar, with the layer spacing larger only by 5% than in the SmY phase [17]. It let us hypothesize that the SmY → Cr2 transition does not affect significantly the relaxation processes, unlike the Cr2 → Cr1 transition, where the structural changes are more significant (14% change in the layer spacing) [17]. The low-frequency process $cr2_{LF}$ has an activation energy of 51.4(3) kJ/mol in the 173-239 K range and 37.8(5) kJ/mol at higher temperatures, where it finally approaches the relaxation time of the $cr1_{LF}$ process from the Cr1 phase upon the Cr2 → Cr1 transition on heating. The high-frequency process $cr2_{HF}$ has an activation energy of only 28(2) kJ/mol. Although the α-relaxation time usually shows non-Arrhenius behavior in organic glassformers, for some compounds the Arrhenius behavior is reported [50,51], corresponding to the low fragility index $m_f \approx 16$. If the $cr2_{LH}$ process is treated as the α-relaxation, then $T_g$ = 145.3(3) K from extrapolation to 100 s. The glass transitions of SmY or Cr2 are not clearly visible in the DSC thermograms, but the rapid slowing down of the cold crystallization with decreasing $T_{cr}$ (Figure 7) indicates that the actual $T_g$ is higher than obtained from the $cr2_{LH}$-relaxation time: the characteristic crystallization time is predicted to reach 24 h at $T_g$ = 226 K, based on the Arrhenius formula fitted for higher temperatures. Thus, the $cr2_{LH}$ process is probably not the α-relaxation.

The potential energy scans for selected torsional angles in the 10OS5 molecule were performed (Figures 10 and 11) for interpretation of the BDS spectra. The $cr2_{HF}$ process may originate either from the simultaneous change of $\varphi_1$ and $\varphi_3$ angles (energy barrier 28 kJ/mol) or $\varphi_2$ and $\varphi_4$ angles (energy barrier 27 kJ/mol), both corresponding to rotation of the COS bridge with one of the phenyl rings. The cr2$_{LF}$ process is related likely to rotation of the phenyl ring via changing of $\varphi_3$ and $\varphi_4$ angles (energy barrier 39.8 kJ/mol). Below 239 K, where $E_a$ is larger, probably both phenyl rings are rotating (summed energy barrier 54.7 kJ/mol). The cr1$_{HF}$ process has an activation energy exceeding significantly the calculated energy barrier for rotation of both phenyl rings. The presence of orientational disorder in Cr1 is unlikely because it was not detected in Cr2, which is metastable and supposed to show a lower order than Cr1. Taking it into account, one can interpret the cr1$_{HF}$ process as originating also from conformation changes, only hindered by stronger intermolecular interactions (e.g. hydrogen bonds) than in the Cr2 phase. The cr1$_{LF}$ process, which does not have an established activation energy, cannot be attributed to any particular intra-molecular rotation, but it can also originate from conformational changes, for the same reason as cr1$_{HF}$.

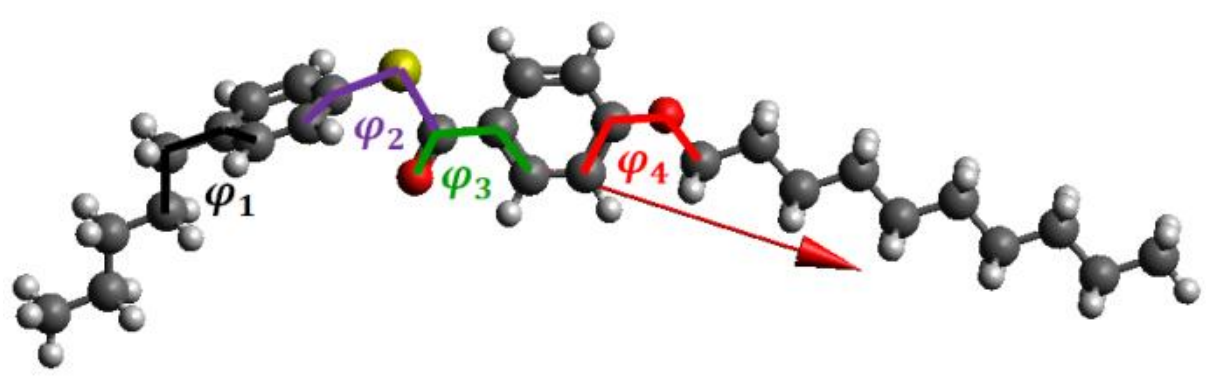

Figure 10. The 10OS5 molecule optimized with the DFT method: B3LYP-D3(BJ) exchange-correlation functional and def2TZVPP basis set ($\varphi_1$ = 272°, $\varphi_2$ = 72°, $\varphi_3$ = 0°, $\varphi_4$ = 180°). The arrow denotes the direction of the total dipole moment (2.7 D).

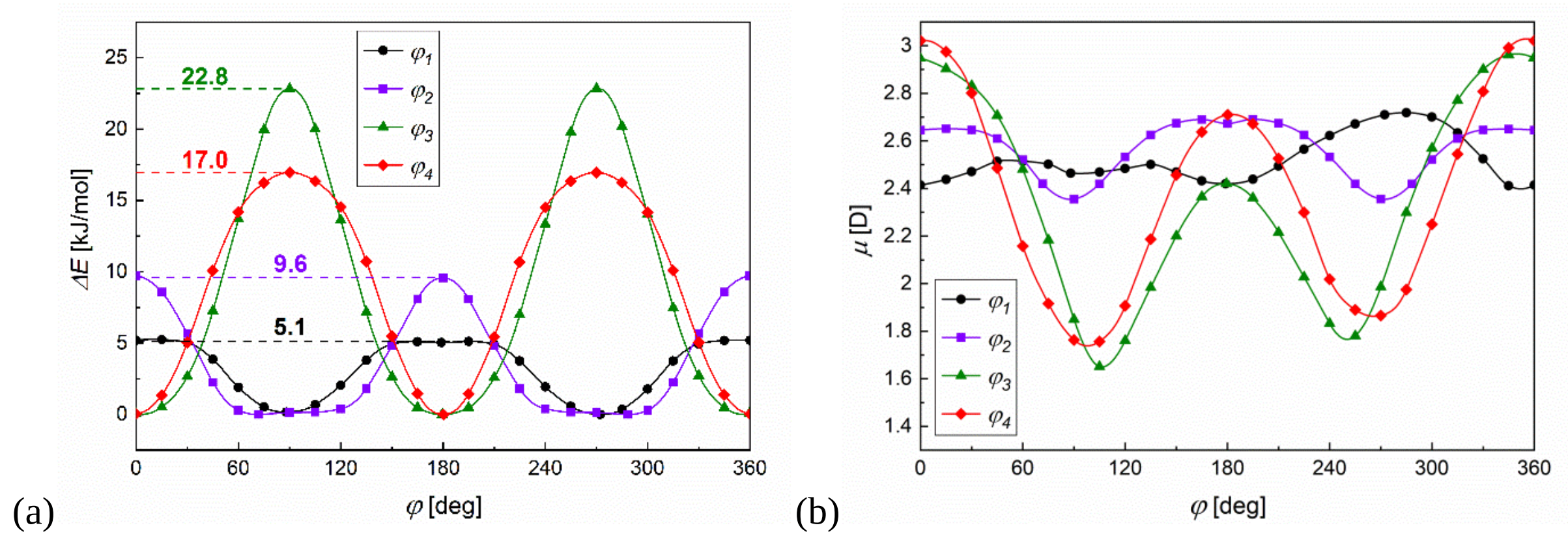

Figure 11. Conformational energy (a) and total dipole moment (b) of the 10OS5 molecule vs. torsional angles indicated in Figure 10.

## 4. Summary and conclusions

The DSC and BDS results of 10OS5 show the sensitivity of the phase sequence on the thermal treatment. Depending on the cooling rate, one can obtain the SmY glass (25-30 K/min), SmY/Cr2 mix (3-20 K/min), or Cr2 (2 K/min). Isothermal crystallization in 290-298 K for a sufficiently long time enables one to obtain the Cr1 phase. Both Cr1 and Cr2, with the melting temperatures 336 K and 331 K, are likely CONDIS phases and they can form glass. The energy release $\Delta H_{cr}$ during cold crystallization on heating increases with the preceding cooling rate, as it is presented in Figure 12a for DSC thermograms with the same cooling/heating rates (Figure 1) and cooling at 30 K/min (Figure 5). The optimal situation for non-isothermal conditions is when the sample is cooled quickly at 30 K/min and heated moderately at 6 K/min, where $\Delta H_{cr}$ = 25.2 kJ/mol. Isothermal cold crystallization leads to larger $\Delta H_{cr}$, which reaches 34.9 kJ/mol at 248 K. Although non-isothermal crystallization occurs in some temperature range, in Figure 12b the $\Delta H$ values are plotted against the peak temperature of the exothermic anomaly, which corresponds to the highest crystallization rate. Application of both slow cooling and heating at 2-6 K/min lead to lower $\Delta H$, but at the same time the region of cold crystallization is shifted to the room temperature and above. Thus, thanks to the metastability of both the SmY and Cr2 phases of 10OS5, one can control, by proper thermal treatment, the temperature range where energy is released. These findings indicate that 10OS5 offers tunable energy release through controlled cold crystallization, making it a promising model system for thermal energy storage based on metastable phases.

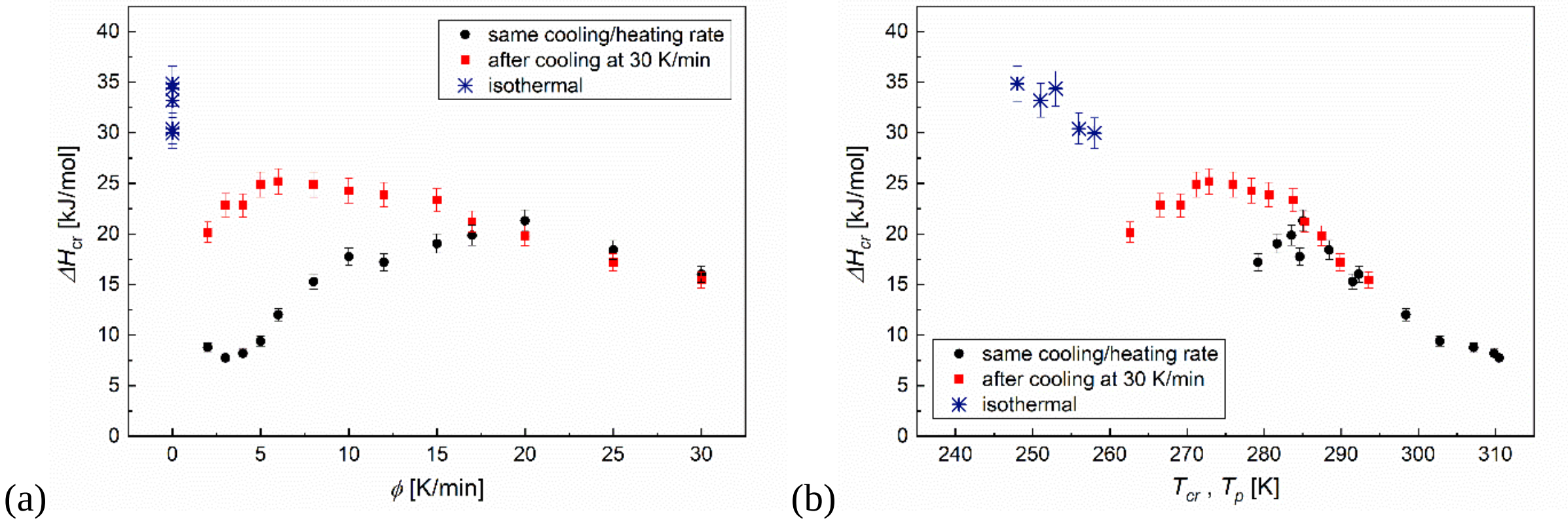

Figure 12. Energy released during cold crystallization of 10OS5 vs. heating rate (c) and temperature (d).

**Acknowledgement:** We gratefully acknowledge the late Assoc. Prof. Małgorzata Jasiurkowska-Delaporte from the Institute of Nuclear Physics, Polish Academy of Sciences for performing the broadband dielectric spectroscopy measurements. We gratefully acknowledge Polish high-performance computing infrastructure PLGrid (HPC Center: ACK Cyfronet AGH) for providing computer facilities and support within computational grant no. PLG/2026/019109.

**Declaration statement:** The authors declare no conflict of interest.